\documentclass[10pt,journal]{IEEEtran}

\usepackage[nocompress]{cite}
\usepackage{subfigure}
\usepackage{multirow}
\usepackage{graphicx}
\usepackage[space]{grffile}
\usepackage{mathtools, nccmath}
\usepackage{dblfloatfix}
\usepackage{amsmath}
\usepackage{dsfont}
\usepackage{wrapfig}
\usepackage[noindentafter]{titlesec}
\usepackage{amsthm}
\usepackage{float}
\usepackage{booktabs}
\usepackage{hyperref}
\usepackage{mathtools}
\usepackage{bbm}
\usepackage{nicefrac}
\usepackage{amssymb,amsfonts}
\usepackage{amsmath}
\usepackage{enumitem}
\usepackage{pifont}
\usepackage{xcolor}
\usepackage{stackengine}

\DeclareMathOperator*{\argmax}{arg\,max}

\newcommand{\blue}[1]{\textcolor{black}{#1}}

\usepackage{bbding}
\usepackage{pifont}
\usepackage{wasysym}
\usepackage{amssymb}
\usepackage{amsthm}
\usepackage[ruled,vlined,noend,linesnumbered]{algorithm2e}
\usepackage{dsfont}

\theoremstyle{definition}

\renewcommand{\paragraph}[1]{\noindent \textbf{#1}}

\begin{document}
\urlstyle{tt}

\title{CILP: Co-simulation based Imitation Learner for Dynamic Resource Provisioning in Cloud Computing Environments}

\author{
        Shreshth~Tuli,
        Giuliano~Casale
    and~Nicholas~R.~Jennings
\IEEEcompsocitemizethanks{
\IEEEcompsocthanksitem S. Tuli, G. Casale are with the Department
of Computing, Imperial College London, United Kingdom.\protect
\IEEEcompsocthanksitem N. R. Jennings is also with Loughborough University, United Kingdom.\protect\\
E-mails: \{s.tuli20, g.casale\}@imperial.ac.uk, n.r.jennings@lboro.ac.uk.\protect
}
\thanks{Manuscript received ---; revised ---.}}


\markboth{IEEE Transactions on Network and Service Management}%
{Tuli \MakeLowercase{\textit{et al.}}: --- }

\maketitle
\thispagestyle{plain}
\pagestyle{plain}

\begin{abstract}
Intelligent Virtual Machine (VM) provisioning is central to cost and resource efficient computation in cloud computing environments. As bootstrapping VMs is time-consuming, a key challenge for latency-critical tasks is to predict future workload demands to provision VMs proactively. However, existing AI-based solutions \blue{tend to not holistically consider} all crucial aspects such as provisioning overheads, heterogeneous VM costs and Quality of Service (QoS) of the cloud system. To address this, we propose a novel method, called CILP, that formulates the VM provisioning problem as two sub-problems of prediction and optimization, where the provisioning plan is optimized based on predicted workload demands. CILP leverages a neural network as a surrogate model to predict future workload demands with a co-simulated digital-twin of the infrastructure to compute QoS scores. We extend the neural network to also act as an imitation learner that dynamically decides the optimal VM provisioning plan. A transformer based neural model reduces training and inference overheads while our novel two-phase decision making loop facilitates in making informed provisioning decisions. Crucially, we address limitations of prior work by including resource utilization, deployment costs and provisioning overheads to inform the provisioning decisions in our imitation learning framework. Experiments with three public benchmarks demonstrate that CILP gives up to 22\% higher resource utilization, 14\% higher QoS scores and 44\% lower execution costs compared to the current online and offline optimization based state-of-the-art methods.
\end{abstract}

\begin{IEEEkeywords}
Resource Provisioning, Cloud Computing, Co-Simulation, Imitation Learning.
\end{IEEEkeywords}

\maketitle

\section{Introduction}\label{sec:introduction}
\IEEEPARstart{T}{he} past years have seen widespread adoption of the cloud computing paradigm due to its flexibility, low maintenance and security. The success of the cloud is attributed to the use of virtualization that allows the deployment of independent virtual machines (VMs) on physical machines (PMs)~\cite{armbrust2010view,le2018cloud}. VMs allow us to efficienty manage the computational resources and reduce operational costs~\cite{zhang2014dynamic}. However, this underpinning technology gives rise to the challenge of efficiently managing resources to ensure optimal service delivery. This becomes crucial for public cloud providers that serve a large number of customers and even companies with private cloud infrastructures to curtail the operational costs of the cloud machines. Resource management is also paramount for users that execute workloads on cloud infrastructures with limited cost budgets~\cite{gu2015cost,chard2015cost,chard2017cost,radhika2021budget}. A key aspect of resource management is VM provisioning, which instantiates and deallocates VMs based on dynamic workload demands. Most prior work aims at the automation of VM provisioning to optimize various performance measures such as energy consumption, operational cost and task response time as we also consider in this work~\cite{narya,al2022deep,ahmad2021dynamic}.

\paragraph{Challenges.} As users shift to workloads that rely on the integration with Internet of Things (IoT) sensors and actuators, contemporary applications have become latency-critical~\cite{schulz2017latency,tuli2022pregan}. In order to make sure that resources are available as soon as possible, proactive VM provisioning is required to avoid degradation of system performance. This entails predicting the resource demands of workloads in a future state and allocating new or deallocating existing VMs for QoS optimization. This broad-level formulation has been widely used in prior work and is commonly referred to as \textit{predictive VM provisioning (PreVMP)}~\cite{uahs,cahs}, which we consider in this work. However, this problem is challenging due to the non-stationary utilization characteristics of most workloads~\cite{ebadifard2021autonomic}, requiring methods to dynamically adapt provisioning policies to the changes in the environment.

\begin{table*}[t]
    \centering
    \caption{Comparison of related works along different parameters (\checkmark means that the corresponding feature is present).}
    \resizebox{\linewidth}{!}{
    \begin{tabular}{@{}lcccccc@{}}
    \toprule 
    \multirow{2}{*}{Work} & \multirow{2}{*}{Method} & \multicolumn{1}{c}{Coupled} & Heterogeneous & Consider & Consider & Low Decision\tabularnewline
     &  & Simulation & Environment & Overheads & Migrations & Time\tabularnewline
    \midrule 
    ARIMA\cite{arima}+ACO\cite{aco} & Autoregressive Prediction + Meta-heuristic Optimization &  &  &  &  & \tabularnewline
    LSTM\cite{lstm}+ACO\cite{aco} & Recurrent NN + Meta-heuristic Optimization &  &  &  &  & \tabularnewline
    Decision-NN\cite{decisionnn} & Deep NN prediction and Gradient based Optimization &  & \checkmark &  &  & \tabularnewline
    Semi-Direct\cite{semidirect} & Dynamic Programming &  & \checkmark &  & \checkmark & \tabularnewline
    UAHS\cite{uahs} & Gaussian Process Regression + Bayesian Optimization &  &  &  &  & \tabularnewline
    Narya\cite{narya} & Deep NN prediction + Multi-Armed Bandits &  & \checkmark &  &  & \checkmark\tabularnewline
    CAHS\cite{cahs} & Gaussian Process Regression + Bayesian Optimization &  &  &  &  & \tabularnewline
    \textbf{CILP} & Imitation Learning + Co-Simulated Oracle & \checkmark & \checkmark & \checkmark & \checkmark & \checkmark\tabularnewline
    \bottomrule
    \end{tabular}}
    \label{tab:related-work}
\end{table*}

\paragraph{Existing solutions.} Several proactive VM provisioning methods have been proposed in the past. Since the optimization objectives, such as cost and response time, depend on \textit{a-priori} unknown future utilization characteristics of workloads, the problem is challenging to solve using conventional optimization strategies. Thus, most state-of-the-art methods decompose the provisioning problem into first predicting the demands of running workloads and then optimizing the VM provisioning plan~\cite{semidirect,uahs,aco}. 
However, existing solutions tend to ignore crucial aspects such as provisioning overheads and heterogeneous VM costs in a cloud system. As VM provisioning entails the time-consuming creation of VMs, this can affect the response time of the new workloads that are placed on new VMs by the underlying scheduler. Moreover, deallocating VMs, in an effort to possibly reduce operation costs, requires the workloads being executed in that VM to be preemptively migrated to other active VMs in the system, further increasing the response times. Higher response times could increase the fraction of the user-defined Service Level Agreement (SLA) violations. Further, cloud providers typically offer a diverse range of VM types with disparate costs and computational capacities. Choosing the optimal set of VMs is crucial to provide the best QoS in the cloud~\cite{sohani2021predictive,zhang2014dynamic}.

\paragraph{Our contributions.} We propose a novel imitation learning framework that leverages a co-simulated digital-twin, \textit{i.e.}, an approximate system model of the cloud infrastructure, to obviate the lack of overhead and cost metrics in decision optimization~\cite{kaur2020convergence,kumar2022efficient}. We term this framework as \underline{C}o-simulation based \underline{I}mitation \underline{L}earner (CILP). Our imitation learner is a composite neural network that predicts future workload demands and provisioning decisions by imitating a co-simulator based decision making oracle. \blue{This allows \textit{CILP} to take optimal provisioning decisions,  ameliorating the need for running costly simulations at test time and enabling \textit{CILP} to efficiently scale with the size of the cloud infrastructure.} Experimental evaluation with three public benchmarks demonstrates that \textit{CILP} gives up to 22\% higher resource utilization, 14\% higher QoS scores, and 44\% lower execution costs than the state-of-the-art techniques.

The rest of the paper is organized as follows. Section~\ref{sec:related} overviews related work.  Section~\ref{sec:formulation} provides the system model assumptions, the co-simulator, underpinning scheduler and the optimization objectives. Section~\ref{sec:approach} presents the CILP provisioner. A performance evaluation of the proposed method on using three public benchmark traces is shown in Section~\ref{sec:experiments}. Finally, Section~\ref{sec:conclusions} concludes and presents future directions.

\section{Related Work}
\label{sec:related}


Dynamic resource provisioning is a long-studied problem in cloud computing~\cite{xu2020dynamic}. When workload demands are static or known, the provisioning task reduces to the classic VM provisioning problem~\cite{cahs}. This has been well studied in the past~\cite{zhao2018power,shahidinejad2021resource}. Several approaches have been proposed that utilize threshold based algorithms~\cite{chard2015cost,gunasekaran2019spock,naha2016cost}, Integer Linear Programming (ILP)~\cite{sharma2010kingfisher} and other estimation based approaches~\cite{han2014enabling}. However, in scenarios with unknown or fluctuating demands, these methods are known to perform poorly~\cite{cahs}. A summary of related work is given in Table~\ref{tab:related-work}.

As previously described, most dynamic resource provisioning methods decouple the provisioning problem into two stages: demand prediction and decision optimization~\cite{uahs}. This is commonly referred to as the \textit{predict+optimize} framework in literature. For the former, a number of methods have been proposed that leverage forecasting models such as AutoARIMA~\cite{arima} or LSTM neural networks~\cite{lstm}. For the latter, conventional methods often use Ant Colony Optimization (ACO)~\cite{aco}, which has been shown to exhibit state-of-the-art QoS scores in recent work~\cite{cahs}. Other methods, such as \textit{Decision-NN}, combine the prediction and optimization steps by modifying the loss function to train neural networks in conjunction with the optimization algorithm~\cite{decisionnn}. This method uses a neural network as a surrogate model to directly predict optimization objectives and uses the concept of neural network inversion, wherein the method evaluates gradients of the objective function with respect to inputs and runs optimization in the input space. The Decision-NN based approach has been shown to be better than gradient-free optimization methods, such as genetic algorithms~\cite{rawat2021resource}. However, continuous relaxation of the discrete optimization problem used in this work has been shown to adversely impact performance~\cite{uahs}. A similar method, \textit{Semi-Direct}, utilizes dynamic programming to find the optimal provisioning decision, but offers limited scalability with workload size~\cite{semidirect}. 

\blue{Recently some provisioning methods have been proposed that utilize deep reinforcement learning to provision VMs in cloud environments~\cite{cheng2018drl,bitsakos2018derp,narya}. \textit{DRL-Cloud} uses deep-Q networks to make provisioning decisions in cloud setups~\cite{cheng2018drl}. \textit{DERP} uses two deep-Q networks to do the same, where having two such networks allows them to reduce bias and improve overall performance of the approach~\cite{bitsakos2018derp}. Another class of methods is referred to as \textit{predictive autoscaling} that scale resources based on demand predictions~\cite{verma2021auto,jindal2019performance}. For instance, the \textit{Autopilot} approach uses sliding windows to identify the CPU/memory limits of individual tasks~\cite{rzadca2020autopilot} and others use repacking to improve system performance~\cite{sedaghat2013virtual}. Similarly, other methods use predictive strategies to take more informed resource allocation and task scheduling decision~\cite{huber2011model}.}

In literature, the current state-of-the-art approaches are \textit{Narya}~\cite{narya}, \textit{UAHS}~\cite{uahs} and \textit{CAHS}~\cite{cahs}. \textit{Narya} is a popular offline approach that is built for mitigating VM interruptions in cloud machines, but can be straightforwardly extended to resource provisioning. It casts the provisioning problem into a Multi-Armed Bandit (MAB) problem, where the objective is to minimize the impact on the QoS of the running workloads. \blue{We use an adapted version of \textit{Narya} as a baseline that uses a neural network as a surrogate model with a MAB model to decide provisioning actions.} \textit{UAHS} is an online optimization approach that leverages a Gaussian Process Regression model for estimating future workload demands and an uncertainty-based heuristic to run Bayesian Optimization over the provisioning decisions. A similar method, \textit{CAHS}, is an extension of \textit{UAHS} that also includes demand correlations while estimating the \textit{utilization ratio} of cloud machines. However, the formulations developed in these works do not consider pragmatic deployment aspects such as provisioning overheads or execution costs while optimizing the provisioning decisions. Further, for large-scale deployments (100+ VMs), these methods tend to get stuck in local optima~\cite{uahs}. As we empirically demonstrate later, these limitations incur performance penalties in cloud environments. We compare \textit{CILP} against \textit{ARIMA+ACO}, \textit{LSTM+ACO}, \textit{Decision-NN}, \textit{Semi-Direct}, \textit{UAHS}, \textit{CAHS} and \textit{Narya} in Section~\ref{sec:experiments}. 

\section{Problem Formulation}
\label{sec:formulation}

\begin{table}[]
    \centering
    \caption{Table of Main Notation}
    \resizebox{\linewidth}{!}{
    \begin{tabular}{@{}ll@{}}
        \toprule
        Notation & Description \\
        \midrule
        $I_t$ & $t$-th scheduling interval \\ 
        $\mathcal{H}_t$ & Set of hosts in interval $I_t$ \\ 
        $\mathcal{W}_t$ & Set of workloads in $I_t$ \\ 
        $c^j_t$ & CPU utilization of workload $w^j_t \in \mathcal{W}_t$ \\
        $r^j_t$ & RAM utilization of workload $w^j_t \in \mathcal{W}_t$ \\
        $s^j_t$ & Disk utilization of workload $w^j_t \in \mathcal{W}_t$ \\
        $W^j_t$ & Feature vectors of workload $w^j_t \in \mathcal{W}_t$ \\
        $\mathcal{A}^i_t$ & Set of workloads allocated to $h^i_t \in \mathcal{H}_t$ \\
        $H^i_t$ & Feature vectors of host $h^i_t \in \mathcal{H}_t$ \\
        $\mathcal{V}$ & Set of VM types \\
        $V^i_t$ & VM type of host $h^i_t \in \mathcal{H}_t$ \\
        $\mu^i$ & Cost per unit time of VM type $V^i_t$ \\
        \blue{$P_t$} & \blue{Provisioning decision for interval $I_t$} \\
        \blue{$D_t$} & \blue{Scheduling decision for interval $I_t$ when  $P_t = \emptyset$} \\
        \blue{$\hat{D}_t$} & \blue{Scheduling decision for interval $I_t$ when $P_t \neq \emptyset$} \\
        $\xi^i$ & Distribution of provisioning time of VM type $V^i_t$ \\ \bottomrule
    \end{tabular}
    }
    \label{tab:symbols}
\end{table}

\subsection{System Model} 
We assume a distributed cloud computing environment with multiple VMs that process a set of independent workloads. We consider a discrete-time control problem, \textit{i.e.}, \blue{we divide the timeline into fixed size execution intervals (of $\Delta$ time duration)} and denote the $t$-th interval by $I_t$. We consider a bounded execution with $T$ intervals; thus, $t \in \{1, \ldots, T\}$. At $I_t$, the set of VM hosts is denoted by $\mathcal{H}_t$ and the set of workloads by $\mathcal{W}_t$. Each workload $w^j_t \in \mathcal{W}_t$ is characterized by its CPU utilization in terms of the number of instructions per second (IPS), denoted by $c^j_t$; RAM utilization in GBs, denoted by $r^j_t$; and disk storage utilization in GBs, denoted by $s^j_t$. Here, $j \in \{1, \ldots, |\mathcal{W}_t|\}$. The feature vector for workload $w^j_t$ is denoted by $W^j_t = [c^j_t, r^j_t, s^j_t]$. The collection of feature vectors of all workloads in $I_t$ is denoted by $W_t$. In $I_t$, we consider a workload allocation, also referred to as a schedule, for each host. The set of workloads allocated to host $h^i_t \in \mathcal{H}_t$ is denoted by $\mathcal{A}^i_t \subseteq \mathcal{W}_t$, where $i\in \{1, \ldots, |\mathcal{H}_t|\}$. 

Similar to workloads, for each host $h^i_t \in \mathcal{H}_t$, the feature vector includes cumulative utilization and maximum capacity of resources (CPU, RAM and disk) and is denoted by $H^i_t = [\sum_{w^j_t \in \mathcal{A}^i_t} W^j_t, \bar{c}^i, \bar{r}^i, \bar{s}^i]$, where $\sum$ denotes vector sum and $\bar{c}^i, \bar{r}^i, \bar{s}^i$ denote IPS, RAM and disk storage capacities of the host. Each host has a VM type that corresponds to a distinct set of utilization capacities and execution costs in a public cloud deployment. We consider a static set of VM types $\mathcal{V}$, where the type of host $h^i_t$ is denoted by $V^i_t = (\bar{c}^i, \bar{r}^i, \bar{s}^i, \mu^i, \xi^i)$. Here, $\mu^i$ is the cost per unit time and $\xi^i$ is the distribution of the provisioning time of a VM instance for a VM type $V^i_t$. We assume $\mu^i$ and $\xi^i$ to be stationary with time $\forall i$. This is common with cloud service providers; for instance, Microsoft Azure charges a constant 0.09 USD per hour for a dual-core \texttt{B2s} machine in its East-US datacenter. Similarly, the provisioning time for a VM type $\xi^i$ remains constant with time as shown by prior work~\cite{mao2012performance} As elements of $V^i_t$ are independent of time, these symbols are not sub-scripted with $t$. $V_t$ denotes VM types for all hosts in $I_t$.  A summary of the symbols is given in Table~\ref{tab:symbols}.

\subsection{VM provisioning} 
\blue{VM provisioning is performed at the start of each interval where we allocate new or deallocate active VMs.} We denote the set of VM provisioning actions at the start of $I_t$ by $P_t$, which is a collection of VM types to be provisioned with the number of instances ($n_t \subseteq \mathcal{V} \times \mathds{Z}$) and hosts to be deallocated from the system ($d_t \subseteq \mathcal{H}_{t-1}$). The new set $\mathcal{H}_t$ is the set of hosts in the previous interval $\mathcal{H}_{t-1}$ union the provisioned hosts in $n_t$ minus the deallocated hosts $d_t$. Deallocation of hosts in the system entails migrating workloads executing in those hosts to other preexisting active or newly provisioned hosts in the system. VM provisioning also entails prediction of the workload utilization characteristics for $I_t$, \textit{i.e.}, $W_t$. We denote this prediction by $\hat{W}_t$, for which the provisioner may use historic data $\{W_k | 1 \leq k < t\}$. We denote the provisioner by $f^{prov}_\theta$ that uses a scheduling decision \blue{$\hat{D}_{t-1}$} and workload utilization metrics of the previous interval $W_{t-1}$ to predict $\hat{W}_t$ and $P_t$. As we use a neural network in our model, we use $\theta$ to denote the weights of such a network. Thus, \blue{$\hat{W}_t, P_t = f^{prov}_\theta(\hat{D}_{t-1}, W_{t-1})$}. We now formulate the underlying scheduler that generates $D_t$.

\subsection{Underlying Scheduler} 
We consider the presence of a scheduler $f^{sched}$ that predicts a schedule $D_t$. \blue{A scheduling decision is the placement of incoming tasks on the set of active VMs in the cloud system.} The scheduler uses the feature vectors of workloads and hosts and a provisioning decision $P_t$. However, the set of workloads $\mathcal{W}_{t-1}$ and hosts $\mathcal{H}_{t-1}$ in $I_{t-1}$ might change in the next interval $I_t$. Thus, the scheduler utilizes $\hat{W}_t$ for existing workloads, $\vec{0}$ for new workloads, $[\vec{0}, \bar{c}^i, \bar{r}^i, \bar{s}^i]$ for new hosts and drops the feature vectors of workloads that complete execution or hosts that are deallocated at the end of $I_{t-1}$ by the provisioner. \blue{If $P_t$ is empty, \textit{i.e.}, it has no provisions or deallocations, then we can use $D_t$ directly to schedule tasks. However, for a non-empty decision $P_t$, the scheduler also needs to decide where to migrate tasks running in hosts that need to be deallocated. In such cases, the tasks running in the hosts that need to be deallocated need to be migrated to other active hosts before deallocating the host. The schedule that also includes these migration decisions is denoted by $\hat{D}_t$.} Thus, with the described modification, the schedule for $I_t$ is evaluated as $\hat{D}_t = f^{sched}(H_{t-1}, \hat{W}_t, P_t)$. The schedule $D_t$ is a bipartite graph with edges from $\mathcal{W}_t$ to $\mathcal{H}_t$ corresponding to the placement of workloads on hosts as in prior work~\cite{tuli2021hunter}. The set $\mathcal{A}^i_t$, described previously, is inferred from $D_t$. The graph nodes are initialized with embeddings corresponding to the feature vectors of workloads and hosts in the system. This graphical modeling of the schedule enables us to scale our neural models with the number of workloads and hosts in the cloud environment. 

\subsection{Co-Simulated Digital-Twin} 

A co-simulated digital twin, referred to as a co-simulator in the rest of the discussion, is a software that models the behavior of a physical system, which in our case is a cloud computing platform. \blue{Several methods in the past have leveraged digital twins of distributed computing environments, such as public clouds, to obviate the need for testing resource management decisions in physical platforms~\cite{borodulin2017towards, hu2018modeling}. Such simulators mimic the behaviors of the physical infrastructure and have been used to generate signals or insights to inform decision making systems~\cite{tuli2021cosco, alam2017c2ps, tuli2022simtune}.} The co-simulator stores the time-series utilization characteristics $W_t$ and $H_t$. It executes a simulation of the cloud model for a given workload features $W_{t-1}$, VM types $V_t$, provisioning decision $P_t$ and scheduling decision \blue{$\hat{D}_t$} to generate QoS metrics for interval $I_t$. These metrics include energy consumption, the response time of completed workloads, SLA violation rates, execution cost and utilization ratio. It also provisions new VMs and deallocates existing ones as per $P_t$.  We denote the co-simulator by $f^{sim}$ and the set of QoS metrics by \blue{$\mathcal{Q}_t = f^{sim}(W_{t-1}, V_t, P_t, \hat{D}_t)$}. 

\subsection{Formulation} 
At the start of the interval $I_t$, given a set of workloads $\mathcal{W}_t$, VM types $\mathcal{V}$, active hosts $\mathcal{H}_{t-1}$ we define our problem as to find a feasible provisioning decision $P_t$ that maximizes the CPU core utilization of hosts to avoid system under-utilization or resource wastage in private clouds. However, this may not capture the heterogeneous pricing policy adopted by cloud providers; thus, we also need to minimize the execution cost for the end user for public cloud deployments. A combination of these two may be required for hybrid cloud environments. We denote these QoS metrics for $I_t$ in our formulation by $r_t$ (utilization ratio) and $\phi_t$ (cost). We define these two metrics as, 
\begin{equation}
\label{eq:obj}
    r_t = \frac{\sum_{w^j_t \in \mathcal{W}_t} c^j_t}{\sum_{h^i_t \in \mathcal{H}_t} \bar{c}^i_t} \text{ and } \phi_t = \sum_{h^i_t \in \mathcal{H}_t} \mu^i \cdot \Delta.
\end{equation}
\blue{The utilization ratio is an important metric that translates to the effective usage efficiency of a cloud system and is a standard metric in prior work~\cite{uahs,cahs}. A higher utilization ratio typically corresponds to lower energy consumption amortized over the number of completed tasks, and thus directly reflects the QoS of the system. Similarly, cost is a crucial metric for both cloud providers and users to reduce the financial footprint of workload execution~\cite{rashid2019cloud}.} These metrics also appear in prior work~\cite{cahs,gu2015cost} and are easy to compute in our formulation, assuming our VM provisioner predicts $c^j_t$s in equation~\eqref{eq:obj}. The formal optimization program is described as follows.
\begin{equation}
\label{eq:problem}
\begin{aligned}
& \underset{\theta}{\text{maximize}} 
& & \sum_{t=1}^T r_t - \gamma \cdot \phi_t \\
& \text{subject to}
& & \forall\ t, \forall\ w^j_t \in \mathcal{W}_t, \sum_{w^j_t \in \mathcal{A}^i_t} W^j_t \leq [\bar{c}^i, \bar{r}^i, \bar{s}^i] \\
&&& \forall\ t, \hat{W}_t, P_t = f^{prov}_\theta(\blue{\hat{D}_{t-1}}, W_{t-1}) \\
&&& \forall\ t, \hat{D}_t = f^{sched}(H_{t-1}, \hat{W}_t, P_t) \\
&&& \forall\ t, r_t, \phi_t \in f^{sim}(W_{t-1}, V_t, P_t, \hat{D}_t) \\
\end{aligned}
\end{equation}
for a given co-simulator $f^{sim}$ and scheduler $f^{sched}$ and $t = \{1, \ldots, T\}$. \blue{In our implementation, we normalize the costs $\mu^i$ by $\max_i \mu^i \cdot \Delta$. Thus, both parts of the convex combination in~\eqref{eq:problem} (i.e., utilization ratio and normalized cost) are in the range [0, 1] and are unit-less quantities.} Further, $\gamma$ is a weighting parameter that can be set by a user based on the deployment scenario and as per the relative importance of the two metrics for the user. Weighting schemes are common to reduce multi-objective optimization to more efficiently solvable single-objective problems~\cite{tuli2022dragon, basu2019learn, tellez2018tabu}. In particular a convex combination allows us to efficiently combine both metrics.

Even with known workload characteristics $W_t$ at the start of $I_t$, the problem is known to be NP-hard~\cite{zhao2018power}.

\begin{figure}
    \centering 
    \includegraphics[width=\linewidth]{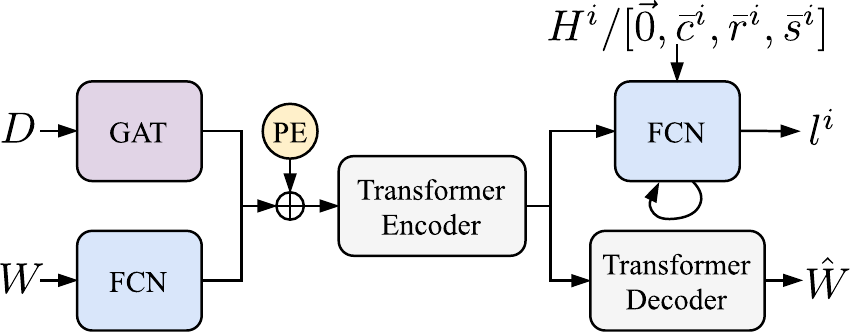}
    \caption{\textit{CILP} neural model encodes the schedule using a GAT and leverages a Transformer to predict utilization characteristics of the next interval and likelihood scores for each provisioning action. }
    \label{fig:model}
\end{figure}

\begin{figure*}[t]
    \centering 
    \includegraphics[width=0.65\linewidth]{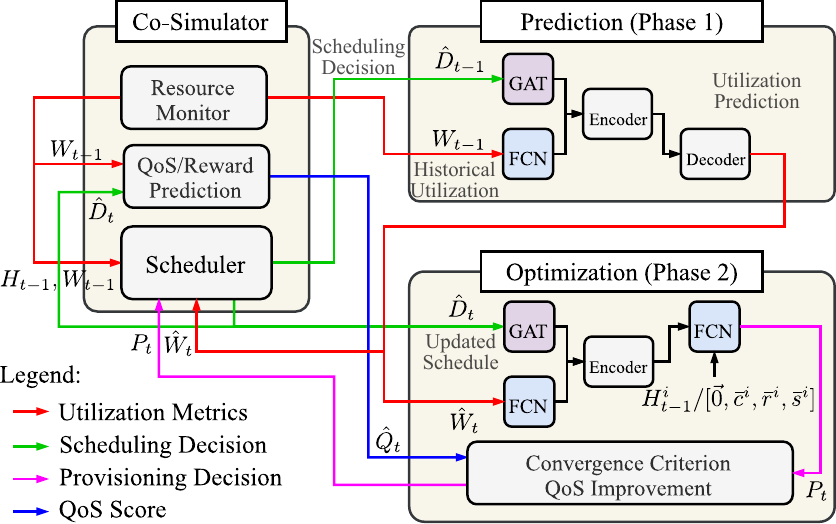}
    \caption{Top level design of the interleaved co-simulation and prediction in \textit{CILP}. For interval $I_t$, with inputs \blue{$\hat{D}_{t-1}$} and $W_{t-1}$, we predict $\hat{W}_t$ and $P_t$ in two phases. In the first phase, the model uses \blue{$\hat{D}_{t-1}$} and $W_{t-1}$ to predict $W_t$. In the second phase the model predicts $l^i_t$'s auto-regressively until the co-simulated QoS score $\hat{Q}_t$ is non-decreasing. $\hat{D}_t$ gets updated after each provisioning action using the scheduler.}
    \label{fig:working}
\end{figure*}

\section{Technical Approach}
\label{sec:approach}

\textit{CILP} learns to predict \textit{a-priori} unknown workload characteristics of the next interval and provisioning decisions and is realized as a neural model $f^{prov}_\theta$. We first describe the neural network based imitation model (Section~\ref{sec:neural-model}), how we infer actions (Section~\ref{sec:inference}), train the model (Section~\ref{sec:training}) and how we translate actions to VM provisions (Section~\ref{sec:provisioning}).

\subsection{Neural Model} 
\label{sec:neural-model}
For interval $I_t$, the inputs of the neural model are schedules \blue{$\hat{D}_{t-1}$} and workload utilizations of the previous interval $W_{t-1}$. To predict $\hat{W}_t$ and $P_t$, we use a composite neural model by inferring the decision using a graph attention network (GAT), the utilization characteristics using a feed-forward network (also referred to as a fully-connected network or FCN) and a Transformer to capture the temporal trends in the data (see Figure~\ref{fig:model} for an overview). An FCN, also referred to as a feed-forward network, is the most basic form of neural network layer that takes an input vector $x$ and uses parameters: weight $W$ and bias $b$ to generate an output
\[y = W \cdot x + b.\]
We stack multiple such layers to create a \textit{deep} neural network. As a result of stacking several of these layers deep neural networks of sufficient capacity are able to approximate functions of arbitrary complexity~\cite{goodfellow2016deep}. In order to achieve this, however, one needs to include non-linear activation functions between the linear layers. We use the $\mathrm{LeakyReLU}$ activation which is shown as
\[\mathrm{LeakyReLU}(x) = \mathds{1}(x \geq 0) \cdot x + \mathds{1}(x < 0) \cdot \epsilon \cdot x,\]
for a small constant $\epsilon$ and $\mathds{1}$ denoting the indicator function. \blue{Unlike the $\mathrm{ReLU}$ activation function that outputs a zero value and has zero gradient for negative inputs, $\mathrm{LeakyReLU}$ gives a non-zero gradient. This allows us to circumvent the dead-neuron problem where the model does not converge to a good optimum~\cite{dubey2019comparative}. Transformers are sequence to sequence models that were initially proposed for NLP. However, in recent literature, they have been shown to outperform recurrent models such as Long-Short-Term-Memory (LSTM) based neural networks particularly because they do not require iterative prediction and allow time-series data to be batched, enabling us to save on training and inference time. Some recent works for resource management in edge and cloud computing environments have shown the promise of Transformers in modeling time-series patterns of workloads and hosts~\cite{xu2022actor,tuli2022simtune,lee2022prediction}.}

For notational convenience, while discussing the neural model, we drop the subscript $t$ without loss in generality. The schedules are encoded as bipartite graph with nodes $\mathcal{N} = \mathcal{W} \cup \mathcal{H}$ and relations $(w^j, h^i) \in D$. Each workload node $w^j$ and host node $h^i$ gets the embeddings $W^j$ and $H^i$ respectively. For a generic node $n \in \mathcal{N}$, we denote its embedding by $e_n$ and its set of neighbors by $\mathcal{S}_n$. Following~\cite{gat}, we perform graph attention convolution as
\begin{align}
\begin{split}
\label{eq:gat}
    \alpha_n &= \mathrm{softmax}_{k \in \mathcal{S}_n} ( \mathrm{LeakyReLU}(W_g^0\ e_k + b_g^0 )),\\
    e_n &= \sigma \big( \sum_{k \in \mathcal{S}_n} \alpha_n \cdot ( W_g^1\ e_k + b_g^1 ) \big),
\end{split}
\end{align}
where $\alpha_n$ are the attention weights for node $n$. \blue{GAT enables the method to efficiently scale with the number of workloads and hosts in the system~\cite{velivckovicgraph}. When we replace GAT with a feed-forward network to infer over the feature vectors of the nodes, the prediction mean-square-error increases by at least 50\% and the inference time by 7\% for the test setups described in Section~\ref{sec:experiments}.} The stacked representation for all nodes ($e_n$) is represented as $E^G$. We also pass the normalized input $W$ of the previous interval through a feed-forward network with sigmoid activation such that
\begin{equation}
\label{eq:window_encoding}
    E^W = \sigma \big(\mathrm{FeedForward}(W) \big).
\end{equation}
We then pass the concatenated vector $E = [E^G, E^W]$ through a Transformer encoder after adding positional encodings (PE)~\cite{vaswani2017attention} using multi-head self-attention, giving an encoded representation,
\begin{equation}
\label{eq:encoder}
    E^0 = \mathrm{TransformerEncoder}(E).
\end{equation}
Early fusion of $E^W$ and $E^G$ enables the downstream predictors to exploit them together. Using $E^0$, we then predict an estimate of utilization characteristics of the next interval, \textit{i.e.} $\hat{W}$, we use a Transformer decoder as
\begin{equation}
\label{eq:decoder}
    \hat{W} = \mathrm{TransformerDecoder}(E^0).
\end{equation}
To generate a provisioning action, we predict the likelihood score for each action independently using a feed-forward network. To allow the model to scale with $|\mathcal{H}|$, for each active host in $\mathcal{H}$, we use the feature vectors from the previous interval $H^i$ and for each new VM type we use the vector $[\vec{0}, \bar{c}^i, \bar{r}^i, \bar{s}^i]$ to infer on which host should be deallocated or provisioned. The likelihood score for each such provisioning action, denoted by $p^i$ with feature vector $F^i$ such that $i \in \{1, \ldots, |\mathcal{H}| + |\mathcal{V}|\}$, is represented as $l^i$ and calculated as
\begin{equation}
\label{eq:fcn}
    l^i = \sigma \big(\mathrm{FeedForward}(E^0, F) \big).
\end{equation}
This factored-style prediction of the likelihood score for each provisioning action enables our model to be agnostic to the number of hosts in the setup. The final provisioning action, denoted as $p$, becomes the action corresponding to the highest likelihood score. These likelihood scores give us a single provisioning action. However, a provisioning decision is a collection of multiple such actions. Moreover, we want to generate likelihood scores using the predicted \blue{workload} utilization metrics $\hat{W}$. To do this, we run inference in two phases. 

\subsection{Two-phase Inference} 
\label{sec:inference}
An overview of the novel two-phase interleaved co-simulation and prediction in \textit{CILP} is presented in Fig.~\ref{fig:working}. We initialize an empty provisioning decision $P_t = \emptyset$. In the \textit{first phase}, for interval $I_t$, using schedule $D_t$ and workload utilization characteristics $W_{t-1}$, we predict $\hat{W_t}$ that estimates the workload demands in $I_t$; thus $\hat{W}_t \gets f^{prov}_\theta(\blue{\hat{D}_{t-1}}, W_{t-1})$. We get $\hat{W}_t$ as the output of the first phase. These prediction estimates of the workload utilization characteristics using the historical values enable the model to make informed decisions for an estimated future system state. Initially, we start without any provisioning; hence, the scheduling decision is obtained as $D_t = f^{sched}(H_{t-1}, W_{t-1}, \emptyset)$. In the \textit{second phase}, we utilize the predicted $\hat{W}_t$ to leverage estimated workload demands in the next interval. Again, we initially keep $P_t$ as an empty set. This gives $\hat{D}_t = f^{sched}(H_{t-1}, \hat{W}_t, \emptyset)$ as the initial schedule. Using $\hat{D}_t$ and $\hat{W}_t$, we evaluate the likelihood scores for each provisioning action $l^i_t = f^{prov}(\hat{D}_t, \hat{W}_t)$. The decided action $p$ is the one with the highest likelihood score. We add this action to $P_t$ and iteratively run the following, updating $P_t$ at each step and evaluating a QoS estimate score, denoted by $\hat{Q}_t$. Thus,
\begin{align}
\label{eq:update}
\begin{split}
    \hat{D}_t &\gets f^{sched}(H_{t-1}, \hat{W}_t, P_t),\\
    l^i_t &\gets f^{prov}(\hat{D}_t, \hat{W}_t),\\
    p &\gets p^{\argmax_i l^i}, \\
    P_t &\gets P_t \cup \{p\},\\
    \hat{Q}_t &\gets f^{sim}(W_{t-1}, V_t, P_t, \hat{D}_t).
\end{split}
\end{align}
Note that the scheduler also decides the preemptive migrations in $\hat{D}_t$ for every non-empty $P_t$ that has host deallocation actions. We continue the above until $\hat{Q}_t$ is non-decreasing. This auto-regressive style of action prediction enables the model to remain parsimonious in terms of the provisioning decisions and avoid excessive overheads. Only those actions are performed that lead to an increase in the expected QoS of the system. Further, the interleaving of action prediction and co-simulation enables us to train an imitation learner, ameliorating the need for costly simulations at test time. Parameter sharing between the demand and likelihood prediction reduces the size of the parameter set, enables \textit{CILP} to jointly learn temporal trends and gain training stability. The converged decision $P_t$ is used for VM provisioning at the start of the interval $I_t$.

\SetKwComment{Comment}{/* }{ */}
\setlength{\textfloatsep}{4pt}
\begin{algorithm}[!t]  
\SetAlgoLined
 \textbf{Require:} Pretrained model $f^{prov}_\theta$, scheduler $f^{sched}$, co-simulator $f^{sim}$\;
 Initialize $W_{-1} \gets \vec{0}$\;
 \For{$t \in \{1, \ldots, T\}$}{
  $\hat{W}_t \gets f^{prov}_\theta(D_t, W_{t-1})$ \Comment*{Predict} \label{line:predict}
  $P_t \gets \emptyset$ \Comment*{Initialize Decision} \label{line:initialize}
  $\hat{D}_t \gets f^{sched}(H_{t-1}, W_{t-1}, \emptyset)$\;
  $l^i_t \gets f^{prov}(\hat{D}_t, \hat{W}_t)$\;
  $p \gets p^{\argmax_i l_i}$\;
  \While{$\hat{Q}_t$ non-decreasing}{
    Update $P_t$, $\hat{Q}_t$ using equation~\eqref{eq:update} \Comment*{Decision Update} \label{line:optimize}
  } 
  Execute $P_t$ and $\hat{D}_t$\;
 }
 \caption{CILP Provisioner} 
 \label{alg:cilp}
\end{algorithm}

\subsection{Model Training} 
\label{sec:training}
We train the \textit{CILP} model using an imitation learning setup where the teacher is the co-simulator, acting as an oracle that generates ground truth actions~\cite{il}. In the second phase, for each input pair $(\hat{D}_t, \hat{W}_t)$, the model generates likelihood score $l^i_t$ for action $p^i$. We also co-simulate the provisioning action $p^i$, generate scheduling decision $\hat{D}_t$ and a reward parameter $\hat{R}_t$ as
\begin{align}
\label{eq:training}
\begin{split}
    \hat{D}_t &\gets f^{sched}(H_{t-1}, \hat{W}_t, \{p^i\}),\\
    r^0_t, \phi^0_t &\gets f^{sim}(W_{t-1}, V_t, \emptyset, D_t),\\
    r^1_t, \phi^1_t &\gets f^{sim}(W_{t-1}, V_t, \{p^i\}, \hat{D}_t),\\
    \hat{R}^k_t &\gets r^k_t - \gamma \cdot \phi^k_t, k \in \{0, 1\},
\end{split}
\end{align}
where $r_t$ and $\phi_t$ are defined as per~\eqref{eq:obj} and superscripts correspond to whether $p^i$ is executed or not (1 if it is). The reward parameter is the objective function in the equation~\eqref{eq:problem}. The ground truth label then becomes $g^i_t = \argmax_k R^k_t$, which signifies whether $p^i$ improves the reward parameter. \blue{Now, for each $p^i$ we evaluate the \textit{imitation loss} as the binary cross-entropy error}
\begin{equation}
    \label{eq:bce}
    L_{BCE}(g^i_t, l^i_t) = - \tfrac{1}{2} \big(g^i \cdot \log(p^i) + (1 - g^i) \cdot \log(1 - p^i) \big).
\end{equation}
We also utilize the mean-square-error between the predicted demands and those from the dataset in the next interval \blue{that we call the \textit{prediction loss}} and is evaluated as
\begin{equation}
    L_{MSE}(\hat{W}_t, W_t) = \frac{1}{|W_t|} \| \hat{W}_t - W_t \|^2.
\end{equation}
Thus, the model training loss for a given utilization trace dataset at each interval $t$ is evaluated as
\begin{equation}
\label{eq:loss}
    L = L_{MSE}(\hat{W}_t, W_t) + \sum_{h^i_t \in \mathcal{H}_t} L_{BCE}(g^i_t, l^i_t).
\end{equation}
Unlike prior work, while calculating the utilization ratio $r_t$, our co-simulator considers the migration delay and uses an average resource utilization value of workloads over the execution interval $I_t$. This makes our optimization objective resemble more closely to real systems.

\subsection{Provisioning in Practice}
\label{sec:provisioning}
Using a pre-trained neural model, the \textit{CILP} provisioner is illustrated in Algorithm~\ref{alg:cilp}. In each interval, we first predict the workload demands in the next interval (line~\ref{line:predict}), initialize decision (line~\ref{line:initialize}) and optimize the preemptive migration decision iteratively as described earlier in equation~\eqref{eq:update} (line~\ref{line:optimize}). The final provisioning decision $P_t$ and schedule $\hat{D}_t$ are then executed in the cloud environment. As shown in the algorithm, each iteration in the CILP training process includes a loop as in equation~\eqref{eq:update} that updates the provisioning decision each time. As a decision could include provisioning a new host, that is one of the VM types $\mathcal{V}$. A decision could also deallocate a VM from the existing set of active hosts in the system, \textit{i.e.}, $\mathcal{H}$. Thus, in the worst case, we run $|\mathcal{H}| + |\mathcal{V}|$ iterations of the inner loop.


\section{Neural Architecture Details}
\label{sec:arch}

We now detail the hyper-parameters for the neural architecture used in \textit{CILP}. We implement all our code using Python-3.8 and PyTorch-1.8.0~\cite{paszke2019pytorch} library.  \blue{All hyper-parameter values are obtained by using grid-search and the \texttt{raytune} library in PyTorch\footnote{\url{https://pytorch.org/tutorials/beginner/hyperparameter_tuning_tutorial.html}.}}. 

\subsection{Graph Attention Network}
The attention weights of the graph attention network (GAT) were evaluated using a feed-forward network of 2-hidden layers, each of size $128$, with $\mathrm{LeakyReLU}$ activation function in all hidden layers. The weights were obtained using a $\mathrm{softmax}$ operation. The embeddings for the nodes of the graph were obtained using graph convolution as a convex combination of neighbor embeddings, with attention weights being used in the combination operation (see equation~\eqref{eq:gat}). The final embeddings ($E^G$) were obtained using the $\mathrm{sigmoid}$ operation. The Parameterized ReLU activation function with a $0.25$ negative input slope was used in the hidden layers for convolution operations. 
%

\subsection{Workload encoder}
We use a feed-forward network with 4-hidden layers, each of size $256$ with $\mathrm{LeakyReLU}$ activation function in all hidden layers. \blue{More layers improve performance; however, for a fair comparison, we ensure that the total parameter count of the neural model in CILP is in the same range as that of the prior work.} The final encoded representation ($E^W$) was obtained using the $\mathrm{sigmoid}$ operation (see equation~\eqref{eq:window_encoding} in Section~\ref{sec:approach}).

%
\subsection{Transformer Encoder }
The input workload utilization characteristics $W_t$ is transformed first into a matrix form of size $|\mathcal{W}_t| \times |W^j_t|$. \blue{We use Transformer encoders and decoders to perform temporal inference over the input workload and host time-series utilization characteristics.} We define scaled-dot product attention~\cite{vaswani2017attention} of three matrices $Q$ (query), $K$ (key) and $V$ (value):
\begin{equation}
    \mathrm{Attention}(Q, K, V) = \mathrm{softmax}\left(\frac{QK^T}{\sqrt{m}}\right)V.
\end{equation}
\blue{For large values of input size (m), the dot product grows large in magnitude, pushing the $\mathrm{softmax}$ function into regions where it has extremely small gradients. To circumvent this, we scale the dot-production attention with $\tfrac{1}{\sqrt{m}}$. } For input matrices $Q$, $K$ and $V$, we apply Multi-Head Self Attention~\cite{vaswani2017attention} by first passing it through $h$ (number of heads) feed-forward layers to get $Q_i$, $K_i$ and $V_i$ for $i \in \{1, \ldots, h\}$, and then applying scaled-dot product attention as
\begin{align}
\begin{split}
    \mathrm{MultiHeadAtt}(Q, K, V) &= \mathrm{Concat}(H_1, \ldots, H_h),\\
    \text{where } H_i &= \mathrm{Attention}(Q_i, K_i, V_i).
\end{split}
\end{align}
Multi-Head Attention allows the model to jointly attend to information from different representation sub-spaces at different positions. In addition, we use position encoding of the input matrices as defined in~\cite{vaswani2017attention}. Our Transformer encoder performs the following operations on the concatenated representation $E = [E^G, E^W]$ as
\begin{align}
\begin{split}
    E^1 &= \mathrm{LayerNorm}(E + \mathrm{MultiHeadAtt}(E, E, E)),\\
    E^0 &= \mathrm{LayerNorm}(E^1 + \mathrm{FeedForward}(E^1)).
\end{split}
\end{align}
Here $\mathrm{LayerNorm}$ is the layer normalization operation described in~\cite{ba2016layer}. We use feed-forward network with 4 layers, each of size $128$ and with $\mathrm{LeakyReLU}$ activation function in all hidden layers. We use number of heads $h = 4$ in our multi-head attention operations. 

%
\subsection{Decoding Predicted Demands} 
Our Transformer decoder generates the predicted workload demands $\hat{W}$ as
\begin{align}
\begin{split}
    E^2 &= \mathrm{LayerNorm}(W_t + E^0),\\
    \hat{W} &= \mathrm{LayerNorm}(E^2 + \mathrm{MultiHeadAtt}(E^2, E^2, E^2)).
\end{split}
\end{align}
Here too, we use a feed-forward network with four layers, each of size $128$ and with $\mathrm{LeakyReLU}$ activation function in all hidden layers and number of heads $h = 4$ in our multi-head attention operations. 

\subsection{Likelihood Prediction}
To predict the likelihood scores, we use the host characteristics $H^i_{t-1}$ of the previous interval or $[\vec{0}, \bar{c}^i, \bar{r}^i, \bar{s}^i]$ for new hosts in the system. We denote each such action as $p^i$ with feature vector $F^i$. Then, $l^i_t$ becomes
\begin{equation}
\label{eq:fcn2}
    l^i = \sigma \big(\mathrm{FeedForward}(E^0, F^i) \big).
\end{equation}
Our feed-forward network consists of 2-hidden layers, each of size $128$, with $\mathrm{LeakyReLU}$ activation function in all hidden layers and $\mathrm{sigmoid}$ activation to generate $l^i$.

\begin{table*}[!t]
    \centering 
    \caption{Comparing utilization ratio $r_t$, cost $\phi_t$ and QoS score $\hat{Q}_t$ for each interval across all models for three public datasets. Values reported include average and standard deviations.}
    \label{tab:results}
    \resizebox{0.95\textwidth}{!}{
    \begin{tabular}{@{}lcccccc@{}}
    \toprule 
    \multirow{2}{*}{Model} & \multicolumn{2}{c}{Azure2017} & \multicolumn{2}{c}{Azure2019} & \multicolumn{2}{c}{Bitbrain}\tabularnewline
    \cmidrule{2-7} 
     & r & cost & r & cost & r & cost\tabularnewline
    \midrule 
    \textit{ARIMA+ACO} & 0.385 $\pm$ 0.103 & 0.481 $\pm$ 0.060 & 0.581 $\pm$ 0.105 & 0.491 $\pm$ 0.056 & 0.374 $\pm$ 0.103 & 0.505 $\pm$ 0.062\tabularnewline
    \textit{LSTM+ACO} & 0.411 $\pm$ 0.124 & 0.508 $\pm$ 0.065 & 0.389 $\pm$ 0.134 & 0.519 $\pm$ 0.059 & 0.401 $\pm$ 0.080 & 0.538 $\pm$ 0.054\tabularnewline
    \textit{Decision-NN} & 0.595 $\pm$ 0.113 & 0.701 $\pm$ 0.022 & 0.732 $\pm$ 0.068 & 0.681 $\pm$ 0.024 & 0.697 $\pm$ 0.081 & 0.647 $\pm$ 0.054\tabularnewline
    \textit{Semi-Direct} & 0.655 $\pm$ 0.065 & 0.679 $\pm$ 0.017 & 0.868 $\pm$ 0.059 & 0.746 $\pm$ 0.057 & 0.733 $\pm$ 0.069 & 0.736 $\pm$ 0.063\tabularnewline
    \textit{UAHS} & 0.753 $\pm$ 0.086 & 0.738 $\pm$ 0.045 & 0.809 $\pm$ 0.060 & 0.701 $\pm$ 0.028 & 0.668 $\pm$ 0.098 & 0.775 $\pm$ 0.038\tabularnewline
    \textit{Narya} & 0.773 $\pm$ 0.068 & 0.607 $\pm$ 0.051 & 0.740 $\pm$ 0.073 & 0.624 $\pm$ 0.032 & 0.696 $\pm$ 0.079 & 0.579 $\pm$ 0.051\tabularnewline
    \textit{CAHS} & 0.800 $\pm$ 0.073 & 0.668 $\pm$ 0.023 & 0.753 $\pm$ 0.065 & 0.779 $\pm$ 0.043 & 0.660 $\pm$ 0.058 & 0.655 $\pm$ 0.040\tabularnewline
    \textit{CILP\_IL} & 0.608 $\pm$ 0.083 & 0.495 $\pm$ 0.025 & 0.736 $\pm$ 0.060 & 0.506 $\pm$ 0.019 & 0.617 $\pm$ 0.106 & 0.492 $\pm$ 0.044\tabularnewline
    \textit{CILP\_Trans} & 0.843 $\pm$ 0.051 & 0.477 $\pm$ 0.016 & 0.817 $\pm$ 0.046 & 0.549 $\pm$ 0.020 & 0.712 $\pm$ 0.056 & 0.441 $\pm$ 0.028\tabularnewline
    \textit{CILP} & \textbf{0.857 $\pm$ 0.049} & \textbf{0.429 $\pm$ 0.020} & \textbf{0.893 $\pm$ 0.051} & \textbf{0.438 $\pm$ 0.036} & \textbf{0.806 $\pm$ 0.051} & \textbf{0.420 $\pm$ 0.029}\tabularnewline
    \midrule 
     & QoS & Training time & QoS & Training time & QoS & Training time\tabularnewline
    \midrule 
    \textit{ARIMA+ACO} & 0.789 $\pm$ 0.008 & 185.212 $\pm$ 0.923 & 0.740 $\pm$ 0.014 & 202.254 $\pm$ 0.317 & 0.749 $\pm$ 0.024 & 361.903 $\pm$ 0.915\tabularnewline
    \textit{LSTM+ACO} & 0.786 $\pm$ 0.008 & 345.824 $\pm$ 0.066 & 0.756 $\pm$ 0.020 & 414.989 $\pm$ 0.281 & 0.759 $\pm$ 0.018 & 657.066 $\pm$ 0.677\tabularnewline
    \textit{Decision-NN} & 0.738 $\pm$ 0.006 & 295.838 $\pm$ 0.411 & 0.736 $\pm$ 0.007 & 355.006 $\pm$ 0.534 & 0.714 $\pm$ 0.021 & 562.092 $\pm$ 0.149\tabularnewline
    \textit{Semi-Direct} & 0.736 $\pm$ 0.005 & 275.894 $\pm$ 0.156 & 0.701 $\pm$ 0.006 & 331.073 $\pm$ 0.827 & 0.711 $\pm$ 0.018 & 524.199 $\pm$ 0.518\tabularnewline
    \textit{UAHS} & 0.715 $\pm$ 0.004 & 315.009 $\pm$ 0.690 & 0.746 $\pm$ 0.008 & 378.011 $\pm$ 0.571 & 0.696 $\pm$ 0.009 & 598.517 $\pm$ 0.449\tabularnewline
    \textit{Narya} & 0.771 $\pm$ 0.003 & 205.196 $\pm$ 0.445 & 0.727 $\pm$ 0.006 & 226.235 $\pm$ 0.940 & 0.727 $\pm$ 0.018 & 399.872 $\pm$ 0.106\tabularnewline
    \textit{CAHS} & 0.738 $\pm$ 0.007 & 281.197 $\pm$ 0.326 & 0.698 $\pm$ 0.013 & 337.436 $\pm$ 0.118 & 0.721 $\pm$ 0.015 & 534.274 $\pm$ 0.757\tabularnewline
    \textit{CILP\_IL} & 0.780 $\pm$ 0.004 & \textbf{133.223 $\pm$ 0.037} & 0.763 $\pm$ 0.007 & \textbf{159.868 $\pm$ 0.595} & 0.736 $\pm$ 0.018 & \textbf{253.124 $\pm$ 0.302}\tabularnewline
    \textit{CILP\_Trans} & 0.796 $\pm$ 0.005 & 291.850 $\pm$ 0.451 & 0.781 $\pm$ 0.010 & 412.220 $\pm$ 0.833 & 0.746 $\pm$ 0.025 & 693.515 $\pm$ 0.841\tabularnewline
    \textit{CILP} & \textbf{0.839 $\pm$ 0.004} & 146.544 $\pm$ 0.789 & \textbf{0.796 $\pm$ 0.011} & 175.853 $\pm$ 0.739 & \textbf{0.803 $\pm$ 0.020} & 278.434 $\pm$ 0.113\tabularnewline
    \bottomrule 
    \end{tabular}}
\end{table*}

\section{Experiments}
\label{sec:experiments}

\label{sec:setup}

\subsection{Datasets}
In order to evaluate the performance of \textit{CILP}, we utilize three public datasets: \texttt{Azure2017}, \texttt{Azure2019} and \texttt{Bitbrain}. The first two are collected from Microsoft Azure public cloud platform and are representative workload traces across thirty consecutive days~\cite{azuredataset}. The \texttt{Azure2017} was generated using 110 cloud VMs, whereas the \texttt{Azure2019} using 150 VMs, both across 30 consecutive days. \blue{The work by Cortez et al.~\cite{azuredataset} identifies the workloads to be a mix of nearly 30\% interactive and 70\% delay-insensitive tasks. The final dataset, \texttt{Bitbrain} consists of traces of resource utilization metrics from 1750 VMs running on BitBrain distributed datacenter~\cite{bitbraindataset}. The workloads running on these servers are from a variety of industry applications including computational analytical programs used by major banks, credit operators and insurers~\cite{bitbraindataset} and are commonly used for benchmarking fog-cloud models~\cite{tuli2021cosco, khamse2018efficient, li2017bayesian}.} We utilize these traces as utilization characteristics of workloads and generate workloads using the same distribution as done in the traces. As all datasets use an interval duration of five minutes, we set the same value as $\Delta$ in our experiments \textit{i.e.} the interval duration in our formulation (see Section~\ref{sec:formulation}). 

\subsection{Baselines} 
We compare \textit{CILP} against 7 baselines. We integrate the \textit{ACO} algorithm with two demand forecasting methods: AutoARIMA and LSTM, and call these \textit{AutoARIMA+ACO} and \textit{LSTM+ACO}. We also include classical predict+optimize methods \textit{Decision-NN} and \textit{Semi-Direct}. Finally, the state-of-the-art baselines are \textit{UAHS}, \textit{Narya} and \textit{CAHS} (see Section~\ref{sec:related}). For a fair comparison, we use the same objective function as given in~\eqref{eq:problem} for all baselines. The implementations of the competitors are not public. We have implemented all baselines based on the model and training details mentioned in the respective papers. For hyperparameters, wherever not mentioned, we use the raytune library\footnote{\url{https://pytorch.org/tutorials/beginner/hyperparameter_tuning_tutorial.html}.} as done for \textit{CILP} (see Section~\ref{sec:arch}). \blue{For a fair comparison, we ensure that the number of parameters in the neural models of each method are similar ($\pm 5\%$ of CILP). Further, we tune the neural networks and hyperparameters of the baselines on the same dataset and with the same tuning approaches as CILP.}

\subsection{Testbed}
We perform our experiments on a Microsoft Azure platform using the Pre-Provisioning Service (PPS) to run our methods, with workloads as \texttt{Docker} containers having utilization characteristics like those of our dataset traces (details in Section~\ref{sec:implementation}). We use diverse VM types in our cloud infrastructure, \textit{i.e.}, \texttt{B2s} with a dual-core CPU and 4GB RAM, \texttt{B4ms} with a quad-core CPU and 16GB RAM and \texttt{B8ms} with an octa-core CPU and 32 GB RAM. All methods may provision up to 200 VMs in our testbed. The $\xi^i$ variables of these VM types, described in Section~\ref{sec:formulation}, are set using Gaussian regression based on 100 datapoints corresponding to the provisioning time for each type. The costs $\mu^i$ are taken from Azure pricing calculator\footnote{\url{https://azure.microsoft.com/en-us/pricing/calculator/}.} for the East-US Azure datacenter~\cite{copeland2015microsoft}. The power consumption values of increments of 10\% CPU utilization of Azure VM types are taken from Standard Performance Evaluation Corporation benchmark repository\footnote{\url{https://www.spec.org/cloud_iaas2018/results/}}.

\subsection{Metrics and Implementation}
\label{sec:implementation}
Our QoS score ($\hat{Q}_t$) is a convex combination of three QoS metrics obtained from the co-simulator, \textit{i.e.}, normalized energy consumption ($q^{e}_t$), average response time ($q^{r}_t$) of completed workloads and SLA violation fraction ($q^{sla}_t$). \blue{For definitions of these metrics, we refer the reader to the COSCO framework~\cite{tuli2021cosco}.} Thus,
\begin{equation}
    \hat{Q}_t \gets 1 - (\alpha \cdot q^{e}_t + \beta \cdot q^{r}_t + \delta \cdot q^{sla}_t),
\end{equation}
where $\alpha, \beta, \delta$ are convex-combination weights set as per \blue{the relative importance of these metrics for the end user. For our experiments, we set them to $\nicefrac{1}{3}$ as per prior work~\cite{tuli2021hunter} for a fair comparison.} The co-simulator $f^{sim}$ obtains energy estimates using simulated power models; it adds migration time and waiting time for provisioned hosts (using $\xi^i$) to the response times of affected workloads. The final response time values are used to decide the violations of SLA deadlines. To implement \textit{CILP},  we build upon the co-simulation primitives provided by the COSCO framework~\cite{tuli2021cosco} by modifying and integrating with custom resource provisioning methods. 
Unlike COSCO, which is for task placement in a \textit{statically provisioned} cloud infrastructure, CILP attacks the much harder problem of provisioning a \textit{dynamic} setup where hosts can be created/destroyed on demand. 
We use the Gradient Optimization using Backpropagation to Input (GOBI) scheduler as our underlying scheduling model and predefined SLA deadlines in COSCO~\cite{tuli2021cosco}. We use the GOBI task scheduler for all baselines as well for a fair comparison. However, unlike GOBI, we do not use a surrogate based QoS predictor for provisioning, but a model that directly predicts decisions. This saves on decision time and the provisioning overhead, which translates to significant improvements in QoS (see Section~\ref{sec:results}). 
 
We use the Pre-Provisioning Service (PPS)~\cite{uahs} in Azure to run our methods with workloads as \texttt{Docker} containers with utilization characteristics as those of traces described in Section~\ref{sec:setup}. Docker is a container management platform as a service used to build, ship and run containers on physical or virtual environments. \blue{In our implementation, we start to provision new VMs at the start of the interval. As soon as a VM has been provisioned, the workload allocation and execution starts.}
 Our containers ran \texttt{sysbench}\footnote{\url{http://manpages.ubuntu.com/manpages/trusty/man1/sysbench.1.html}.} and \texttt{iozone}\footnote{\url{https://linux.die.net/man/1/iozone}.} linux benchmarking tools. The former facilitates matching the IPS of the utilization traces of the datasets and the latter matches the RAM and storage consumption.

\blue{We run the CILP provisioner (including the imitation model and co-simulated digital-twin) and scheduler on a broker machine that manages the set of Azure workers. The broker node has the following configuration: Intel i7-10700K CPU, 64GB RAM, Nvidia RTX 3080 and Windows 11 OS. The collection of dataset and training of the CILP model was performed on the same machine.}

We use HTTP RESTful APIs for communication and seamless integration of a \texttt{Flask} based web-environment to deploy and manage containers in our distributed cloud setup~\cite{grinberg2018flask}. For preemptive migrations, we use the Checkpoint/Restore In Userspace (CRIU)~\cite{venkatesh2019fast} tool for container migration. All sharing of resource utilization characteristics across workers uses the \texttt{rsync}\footnote{\url{https://linux.die.net/man/1/rsync}.} utility. For synchronization of outputs and execution of workloads, we utilize the HTTP Notification API.

\begin{figure*}[!t]
    \centering
    \includegraphics[width=\linewidth]{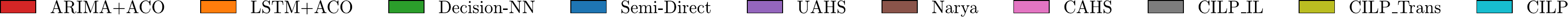} \\
    \subfigure[Energy Consumption]{
    \includegraphics[height=.195\textwidth]{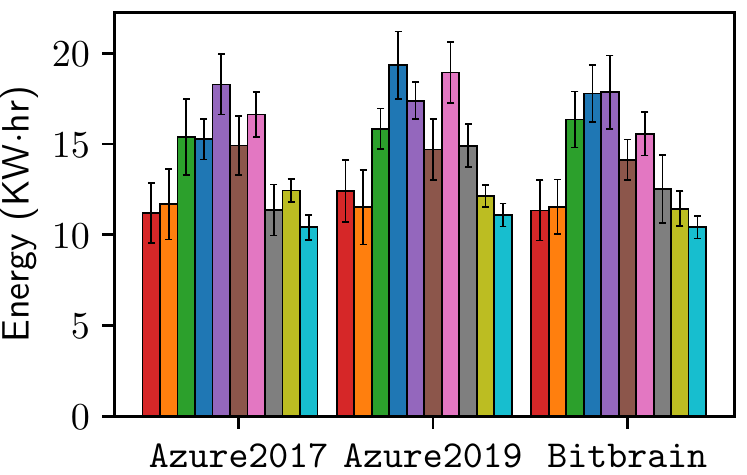}
    \label{fig:energy}
    }
    \subfigure[Response Time]{
    \includegraphics[height=.195\textwidth]{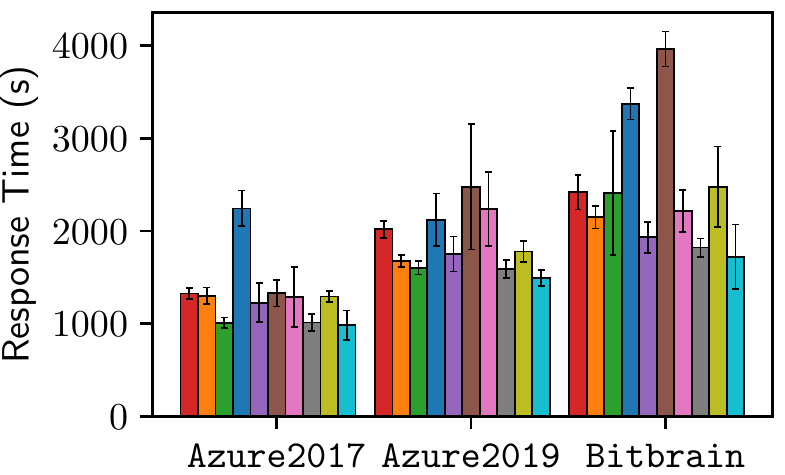}
    \label{fig:response}
    }
    \subfigure[SLA Violation Rate]{
    \includegraphics[height=.195\textwidth]{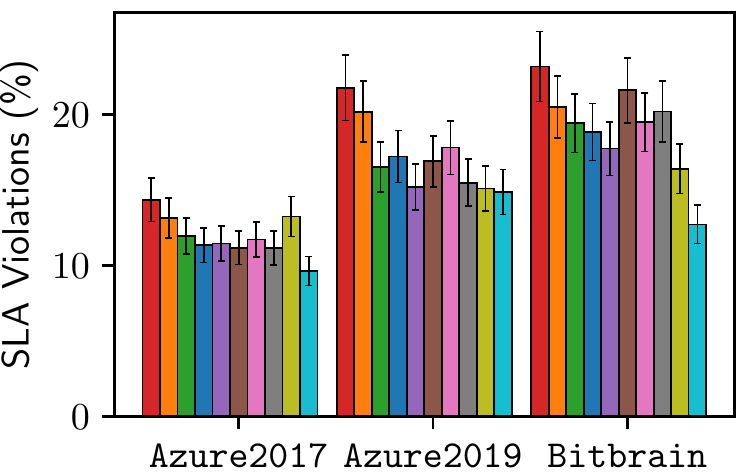}
    \label{fig:sla}
    }\\
    \subfigure[Waiting Time]{
    \includegraphics[height=.195\textwidth]{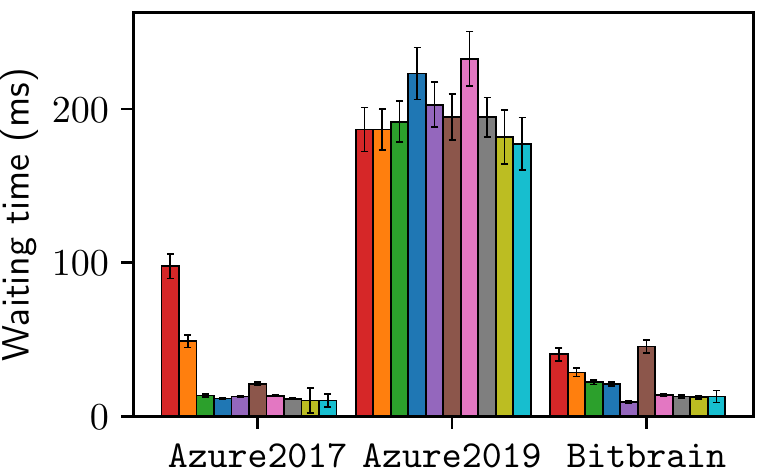}
    \label{fig:waiting}
    }
    \subfigure[Provisioning Overhead]{
    \includegraphics[height=.195\textwidth]{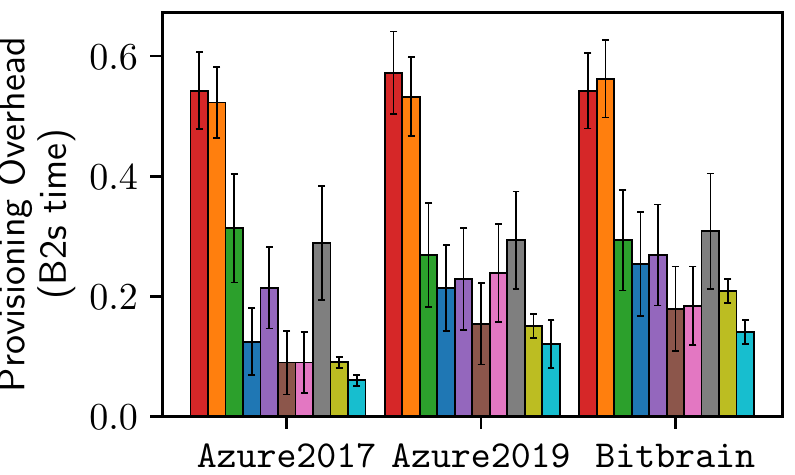}
    \label{fig:overhead}
    }
    \subfigure[Migration Count]{
    \includegraphics[height=.195\textwidth]{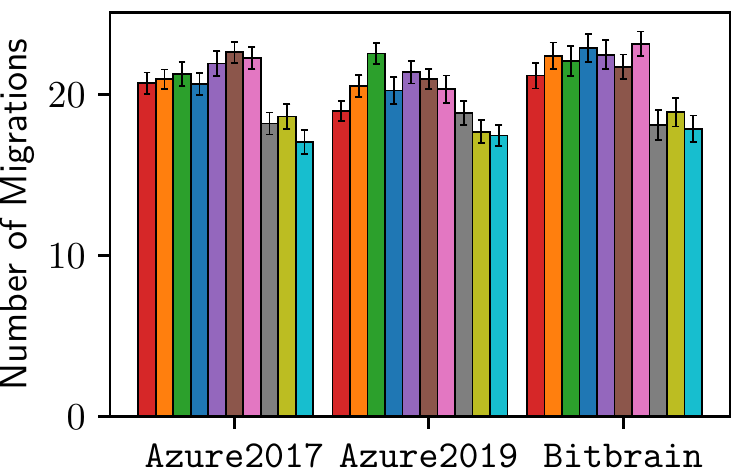}
    \label{fig:migrations}
    }
    \caption{Comparison of QoS parameters (averaged over intervals) of \textit{CILP} against baselines and ablated models. }
    \label{fig:results}
\end{figure*}

\subsection{Experimental Details}
\label{sec:experimental_details}
\blue{The \textit{CILP} and baseline methods run on a dedicated broker node as described previously.} We run for $T = 200$ intervals. We set the user-defined parameter $\gamma$ using grid-search over the average QoS score $\tfrac{1}{T} \sum_{t=1}^T \hat{Q}_t$ generated from the co-simulator (see sensitivity analysis in Section~\ref{sec:results}). The neural network architecture used for the experiments is detailed in Section~\ref{sec:arch}. 

\subsection{Training Details}
Model training used a learning rate of $5\times 10^{-4}$. The Adam optimizer~\cite{kingma2014adam} with a weight decay parameter of $10^{-5}$ and a batch size of $64$ was used. As described in Section~\ref{sec:experimental_details}, we use data corresponding to 200 intervals from the datasets \texttt{Azure2017}, \texttt{Azure2019} and \texttt{Bitbrain} to test the model. The rest is used to generate the training data to generate the ground-truth utilization characteristics and likelihood scores of provisioning actions using the co-simulated oracle. To train the neural model, we randomly divide the training time series into $80\%$ training data and $20\%$ validation data. \blue{We observe that temporal correlations in the training data are often short-range and mid-range in the setting we consider. Hence, we sample 200 points in a minibatch - since the autocorrelation function becomes flat after lag 50. This is a conservative sample length that ensures that correlations with significant magnitude are not damaging the temporal structure of the time series.} An early stopping criterion was applied for convergence using the value of the loss function of the validation data. 

\blue{The loss value $L$ given by equation~\ref{eq:loss} for the converged model after the training process was $2 \times 10^{-3}$. The prediction performance of the CILP neural model directly affects the effective QoS of the system. In case the MSE error of predicting the workload utilization characteristics ($\hat{W}_t$) is high in phase 1, the scheduler would not be able to take well informed scheduling decisions to allocate incoming workloads or running tasks from hosts that need to be deallocated. Further, in case of a high BCE loss in equation~\ref{eq:bce}, the model poorly imitates the oracle decisions from the co-simulator and thereafter takes poor provisioning decisions. Thus, it is critical for both prediction and imitation losses to be low for an effective model.}

\subsection{Comparison with Baselines} 
\label{sec:results}
Table~\ref{tab:results} presents the utilization ratio, cost in USD and QoS score ($\hat{Q}_t$) averaged over the execution intervals, and training times of all models. Figure~\ref{fig:results} shows the individual QoS metrics for each method. For all datasets, \textit{CILP} outperforms the baselines in terms of average utilization ratio, execution cost and QoS scores. Compared to state-of-the-art baselines, \textit{i.e.}, \textit{UAHS}, \textit{Narya} and \textit{CAHS}, \textit{CILP} gives up to 13.81\%-22.12\% higher values of average utilization ratio, 41.87\%-45.80\% lower average execution cost and 14.04\%-17.34\% higher average QoS score. We also observe that \textit{CILP} gives lower training times compared to all baselines. \textit{CILP} gives  30.31\%-57.68\%. This clearly demonstrates the advantage of having a transformer model with positional encoding to push the complete time-series data as an input instead of sequentially inferring over local windows as in auto-regressive (\textit{ARIMA}), recurrent models (\textit{LSTM}) or feed-forward models (\textit{Decision-NN}, \textit{Semi-Direct}, \textit{UAHS}, \textit{Narya}, \textit{CAHS}).

When comparing individual QoS metrics (see Figure~\ref{fig:results}), \textit{CILP} gives the least average energy consumption of 10.41 KW$\cdot$hr, 7.78\% lower than \textit{ARIMA+ACO} with lowest energy consumption across all baselines. This is due to the cost minimization in \textit{CILP} that ensures the least number of hosts are active in the system, leading to a lower energy footprint. \textit{CILP} also gives the lowest average response time, 7.41\% lower than the best baseline \textit{Decision-NN} that leads to the lowest SLA violation rates that are up to 28.32\% lower than the best baseline \textit{UAHS}. This is due to the QoS augmented decision strategy in the \textit{CILP} methodology. Additionally, the improvements in response times come directly from proactive provisioning in \textit{CILP}, which minimizes the time tasks spent waiting to be scheduled, as shown in Figure~\ref{fig:waiting}. \blue{Finally, \textit{CILP} also gives the least overheads in terms of provisioning delays and causes the minimum number of migrations, thanks to its co-simulation driven stopping criterion in Algorithm~\ref{alg:cilp}.}

\subsection{Ablation Analysis} 
To test the efficacy of the co-simulated imitation learning and Transformer based neural models in \textit{CILP}, we modify the approach as follows. First, we consider a model without co-simulated reward scores $\hat{R}_t$, but instead evaluate equation~\eqref{eq:obj} without migration and provisioning overheads to train the neural model. We call this the \textit{CILP\_IL} model. Second, we replace the Transformer encoder and decoder with feed-forward networks (with the same number of parameter weights) to test the importance of temporal trends that the Transformer captures. We call this the \textit{CILP\_Trans} model. The results in Table~\ref{tab:results} and Figure~\ref{fig:results} show a drop in all performance metrics for these models when compared to \textit{CILP}, demonstrating the effectiveness of the Transformer based neural architecture and model training using co-simulated imitation learning. Even though the training times of the \textit{CILP\_IL} model is the lowest, the lack of overhead information does not allow it to take informed decisions, giving rise to poor QoS scores.

\begin{figure*}
    \centering
    \subfigure[\texttt{Azure2017}]{
    \includegraphics[height=.135\textwidth]{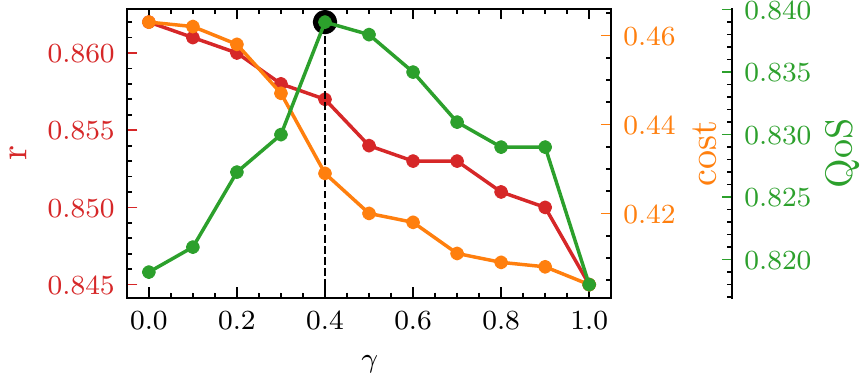}
    \label{fig:sens_azure17}
    }
    \subfigure[\texttt{Azure2019}]{
    \includegraphics[height=.135\textwidth]{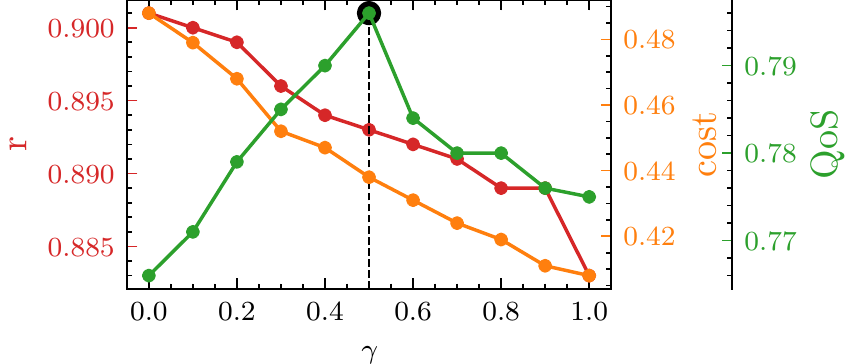}
    \label{fig:sens_azure19}
    }
    \subfigure[\texttt{Bitbrain}]{
    \includegraphics[height=.135\textwidth]{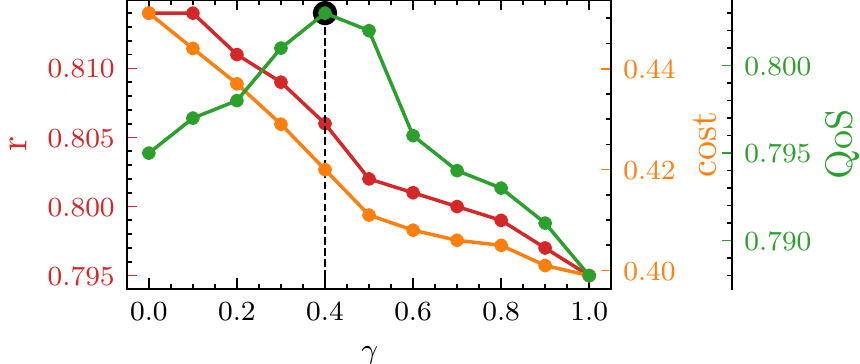}
    \label{fig:sens_bitbrain}
    }
    \caption{Sensitivity analysis with the weight of the execution cost in the objective function ($\gamma$). All standard deviations are $<$0.05. The $\gamma$ values that give the highest QoS have been highlighted. }
    \label{fig:sens}
\end{figure*}

\subsection{Sensitivity Analysis}
Figure~\ref{fig:sens} shows the performance of the \textit{CILP} method for different values of  user-defined hyper-parameter $\gamma$. Low $\gamma$ corresponds to private-cloud deployments that do not have any monetary running costs, but require maximization of utilization ratio to reduce resource wastage. High $\gamma$ corresponds to workload execution on public clouds with limited cost budgets, where the user aims at optimizing both system utilization and running costs. For comparison with baselines, we use $\gamma = 0.4, 0.5, 0.4$ for the three datasets, as per grid-search, while maximizing QoS scores. The \textit{CILP} provisioner gives higher QoS scores than each baseline for every value of $\gamma$. These results provide clear evidence of the robustness of the \textit{CILP} to different deployment scenarios and hyper-parameter settings.

\subsection{Discussion}

The results demonstrate that \textit{CILP} outperforms baselines not only in terms of QoS parameters, but also gives lower overheads. As provisioning overhead is hard to predict in dynamic settings, it is technically challenging to include it as part of the optimization objective. Unlike prior methods that run optimization using a model (\textit{ARIMA+ACO}, \textit{Decision-NN}, \textit{Semi-Direct}, \textit{Narya}, \textit{UAHS} and \textit{CAHS}), \textit{CILP} directly generates an action, saving search time and, consequently, the provisioning overhead. A higher provisioning overhead leads to increased decision time and slower availability of required resources. \textit{CILP}’s lower provisioning overhead reduces the average response time and SLA violation rates, significantly improving the system QoS.

Akin to a \textit{CILP} style imitation learner, our adapted version of \textit{Narya} is the only baseline that learns to predict provisioning actions directly (using MAB) instead of running online optimization of the provisioning decision at run-time (such as other baselines \textit{LSTM+ACO}, \textit{Decision-NN}, \textit{SemiDirect}, etc.). Unlike online-optimization based baselines, the adapted \textit{Narya} model gives lower provisioning overheads, as seen in Figure~\ref{fig:overhead}. However, the use of co-simulation based stopping criterion in the two-phase loop (line~\ref{line:optimize} in Algorithm~\ref{alg:cilp}) leads to lower number of VM provisions and consequently reduced provisioning overheads. Our results corroborates that \textit{CILP} outperforms even offline learners such as \textit{Narya} and as well as online optimization based provisioning methods. 

\section{Conclusions}
\label{sec:conclusions}

This work fundamentally focuses on the VM provisioning problem. We formulate it as a predict+optimize problem. We propose \textit{CILP}, a co-simulated digital-twin based imitation learner that dynamically predicts VM provisioning decisions to optimize the QoS of a cloud computing environment. It uses a Transformer based composite neural network to first predict the workload demands in the near future and auto-regressively find the optimal set of provisioning actions. Imitation learning enables us to train the action predictor in order to imitate the QoS effects of each action using a co-simulator. 

\blue{As we discuss related works, we demonstrate that prior methods optimize primarily the utilization ratio as a metric, and do not consider the counter-metric of the operational cost of the system. The proposed CILP approach uses cost as an optimization metric explicitly in its objective function. To make the final provisioning decisions, we use the QoS estimate from our co-simulation model, which uses a convex combination of the real QoS metrics: energy consumption, response time and SLA violation rate. Using this simulator, we are able to implicitly identify the impact of various provisioning decisions on the real QoS metrics, including the impact of overheads of VM provisioning (in the form of higher response time, for instance) that is modeled in the simulator.} Extensive experiments on three public datasets show that \textit{CILP} outperforms the state-of-the-art in QoS metrics, including energy consumption, resource utilization, execution cost and SLA violation rates. 

As part of future work, we shall explore how \textit{CILP} can be extended to not only make optimal provisioning decisions, but also workload scheduling decisions using the co-simulator. The predict+optimize formulation is applicable to other problem settings as well, allowing us to leverage \textit{CILP} for decision making wherever there is access to a dedicated infrastructure and its co-simulator.

\section*{Software Availability}
The code has been made available as a public GitHub repository under BSD-3 License at \url{https://github.com/imperial-qore/CILP}. 

\section*{Acknowledgments}
Shreshth Tuli is supported by the President's PhD scholarship at Imperial College London. 

\bibliographystyle{IEEEtran}
\bibliography{ijcai22}

\begin{IEEEbiography}
[{\includegraphics[width=1in,height=1in,clip,keepaspectratio]{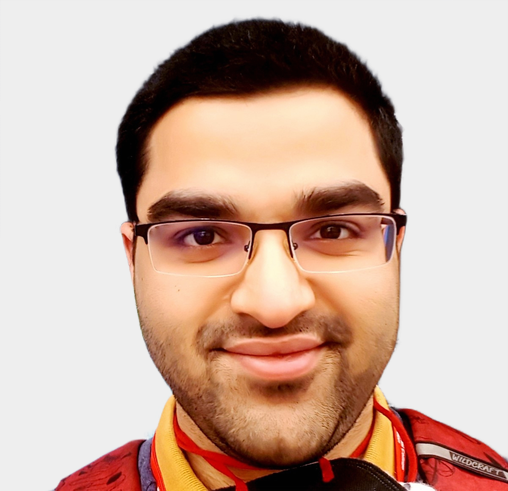}}]
{Shreshth Tuli}
is a President's Ph.D. Scholar at the Department of Computing, Imperial College London, UK. Prior to this he was an undergraduate student at the Department of Computer Science and Engineering at Indian Institute of Technology - Delhi, India. He has worked as a visiting research fellow at the CLOUDS Laboratory, School of Computing and Information Systems, the University of Melbourne, Australia. He is a national level Kishore Vaigyanik Protsahan Yojana (KVPY) scholarship holder from the Government of India for excellence in science and innovation. His research interests include Fog Computing and Deep Learning. For further information, visit \url{https://shreshthtuli.github.io/}.
\end{IEEEbiography}
\begin{IEEEbiography}
[{\includegraphics[width=1in,height=1.25in,clip,keepaspectratio]{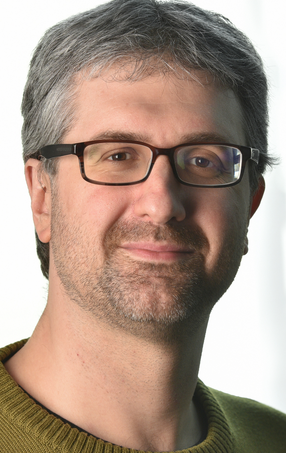}}]
{Giuliano Casale}
joined the Department of Computing at Imperial College London in 2010, where he is currently a Reader. He teaches and does research in performance engineering and cloud computing, topics on which he has published more than 150 refereed papers. He has served on the technical program committee of several conferences in the area of performance and dependability. His research work has received multiple recognitions, including best paper awards at ACM SIGMETRICS, IEEE/IFIP DSN, and IEEE INFOCOM. He serves on the editorial boards of IEEE TNSM and ACM TOMPECS, as editor in chief of Elsevier PEVA, and as the current chair of ACM SIGMETRICS.

\end{IEEEbiography}
\begin{IEEEbiography}
[{\includegraphics[width=1in,height=1.25in,clip,keepaspectratio]{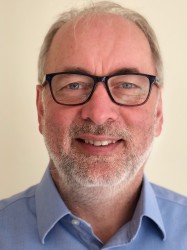}}]
{Nicholas R. Jennings}
is the Vice-Chancellor and President of Loughborough University. He is an internationally-recognised authority in the areas of AI, autonomous systems, cyber-security and agent-based computing. He is a member of the UK government’s AI Council, the governing body of the Engineering and Physical Sciences Research Council, and chair of the Royal Academy of Engineering’s Policy Committee.  Before Loughborough, he was the Vice-Provost for Research and Enterprise and Professor of Artificial Intelligence at Imperial College London, the UK's first Regius Professor of Computer Science (a post bestowed by the monarch to recognise exceptionally high quality research) and the UK Government’s first Chief Scientific Advisor for National Security.
\end{IEEEbiography}

\vfill

\end{document}